\documentclass[
showpacs,preprintnumbers,amsmath,amssymb,
prb]{revtex4}
\usepackage{graphicx}
\usepackage{dcolumn}
\usepackage{bm}
\usepackage{amssymb}
\usepackage{amsmath}
\usepackage{epsf}

\begin{document}
\title{Slowing down of  spin relaxation in two dimensional
systems by quantum interference effects.}

\author{I.~S.~Lyubinskiy,  V.~Yu.~Kachorovskii}
\affiliation{A.F.~Ioffe Physical-Technical Institute, 26
Polytechnicheskaya str., Saint Petersburg, 194021, Russia}

\date{\today}
\begin{abstract}
{ The effect of weak localization  on  spin relaxation in a two-dimensional system with a
spin-split spectrum is considered. It is shown that the spin relaxation slows down due to the interference of
electron waves moving along closed paths in opposite directions. As a result, the averaged electron
spin
 decays at large times as $1/t$. It is found that the  spin dynamics
 can be described
 by a Boltzmann-type  equation, in which  the
   weak localization effects  are taken into account as  nonlocal-in-time corrections
   to  the collision integral. The corrections are
   expressed via a
 spin-dependent return probability.
 The physical nature of
the phenomenon is discussed and it is shown  that the "nonbackscattering" contribution to the weak localization plays
an essential role.  It is also demonstrated  that the magnetic field, both transversal and longitudinal,
suppresses the power   tail in the spin polarization. }
\end{abstract}
 \maketitle

\section*{Introduction}
  The relaxation  of
    non-equilibrium spin
 polarization is  the central phenomenon in
 spin-dependent  transport   in semiconductor
 nanostructures. \cite{avsh}
 One of the most efficient mechanisms of  electron spin relaxation in III-V semiconductors is the
well-known Dyakonov-Perel mechanism \cite{perel} based on the classical picture of  angular
diffusion of the  spin vector in a random magnetic field. The field originates from the
momentum-dependent  spin-orbit splitting of the conduction band in the crystals with zinc-blend
structure. \cite{dress} While passing through the crystal, the electron is scattered by
impurities and its momentum changes randomly with time. As a consequence, the effective magnetic
field also changes randomly with a correlation time of the order of the momentum relaxation time
$\tau$.
 The spin
 relaxation time $\tau_S$ is  a characteristic time of the spin
angular diffusion \cite{perel} $~~1/\tau_{S}\sim \phi^2/\tau \sim \Omega^2\tau$ , where $\phi \sim
\Omega\tau \ll 1$ is the typical angle of the spin precession for the momentum relaxation time and
$\Omega$ is the frequency of precession in the effective magnetic field proportional to the
conduction band splitting.  For a two-dimensional (2D) case, when the electron motion in one direction is confined by
the quantum well, the spin splitting and, hence, the precession frequency are proportional to the
 in-plane electron velocity \cite{rashba,dyak}  $~\Omega\sim v.$ So, at low temperatures, when
inelastic processes can be disregarded,   the spin relaxation rate for a 2D electron with a given
energy $E=mv^2/2$ is proportional to the particle diffusion
coefficient
\begin{equation}
\frac{1}{\tau_{S}}\sim D,\qquad D=v^2\tau/2.
 \label{tau}
\end{equation}
The  effects of localization on the particle diffusion  have been discussed in a great number of
publications. The first-order term in a series expansion of $D$  in $1/kl$ ($l=v\tau$ is the mean
free path, $k$ is the electron wave vector)   is known as the  weak localization correction
\cite{wl} ( for a review, see Ref.~\onlinecite{lee}), coming from the coherent enhancement of the
backscattering amplitude. A remarkable feature of this correction is the logarithmic divergence at
low temperatures in the 2D case. Equation~\eqref{tau} implies a similar divergence of the spin relaxation
rate. Such a divergence was first predicted by Singh \cite{sin1} in spin correlation functions for
a system with spin-dependent impurity scattering. It was shown, however, that the quantum
correction to the spin relaxation rate is not proportional to the quantum correction to the
diffusion coefficient, as one might  expect from Eq.~\eqref{tau}. A similar result    was obtained
in Ref.~\onlinecite{m1} for a system with a spin-split spectrum.
 It was found in Ref.~\onlinecite{m2}  that the weak localization slows down the spin relaxation
of excitons in quantum wells, which leads to a $1/t$ power tail in the spin orientation.   A similar
  effect was also
predicted for
electrons in 2D semiconductors with a zinc-blend crystal structure.

In this paper,  we consider the  effects of localization on the spin relaxation for a 2D semiconductor
with a spin-split spectrum. We show that the spin dynamics is described by a Boltzmann-type
equation. In the first order in $1/kl,$ the localization effects can be taken into account by a
nonlocal-in-time correction to the Boltzmann collision integral. This correction is expressed in
terms of the spin-dependent return probability. We discuss the role of coherent returns at
different scattering angles and show that the "nonbackscattering" contribution to the collision
integral plays a key role.  We solve  the generalized kinetic equation and demonstrate that, at
large times, the spin polarization decays as $1/t.$ The magnetic field, both transversal and
longitudinal, is found to suppress the long-living tail in the spin-relaxation.

\section*{Derivation of the kinetic equation}

The Hamiltonian of a 2D with a spin-split spectrum is given by
\begin{equation}
H=\frac{\mathbf p^2}{2m}+\frac{\hbar}{2}\boldsymbol{\omega}\boldsymbol{\sigma} +U(\mathbf
r).
\label{hamilt}
\end{equation}
Here $\mathbf p=p\mathbf n$ is the  in-plane electron momentum, $m$ is the electron effective mass
and $\boldsymbol {\sigma}$ is a vector consisting of  Pauli matrices. The spin-orbit interaction
is described by the second term, in which $\boldsymbol{\omega}=\boldsymbol{\omega}(\mathbf n)$ depends
on the direction of the electron momentum $\omega_i(\mathbf n)= \sum_k n_k\Omega_{ki} ~~(i=x,y,z;
~k=x,y) .$  The matrix $\hat\Omega =
 \hat \Omega^{(1)} + \hat \Omega ^{(2)} $  is  the sum of two terms:
  the so-called Bychkov-Rashba term \cite{rashba} $\hat \Omega^{(1)}$ (with nonzero components
  $\Omega^{(1)}_{xy}=-\Omega^{(1)}_{yx}~\sim p $)
 and $\hat \Omega^{(2)},$ which  is the Dresselhaus
term \cite{dress} averaged  over the electron motion along the $ z$-direction perpendicular to the
quantum well plane. The Bychkov-Rashba coupling depends on the asymmetry of the quantum well
confining potential. Its  strength  can be tuned by varying the gate voltage.\cite{nitta} The Dresselhaus
term is present in semiconductors with no bulk inversion symmetry. The components of the matrix
$\hat \Omega^{(2)}$ are also linear in the  in-plane electron momentum $ \Omega^{(2)}_{ij} \sim p$
and vary with well plane   orientation
   with respect to  the crystallographic axes \cite{dyak} (we neglect cubic Dresselhaus terms,
   assuming that the electron concentration is relatively small).
   We consider the scattering by the short-range impurity potential with the correlation function
$\langle U(\mathbf r) U(\mathbf r')\rangle = \gamma \delta (\mathbf r -\mathbf r')$, where the
coefficient  $\gamma$ is related to the transport scattering time by  $\tau=
\hbar^3/m\gamma .$

The classical spin dynamics   is described by the kinetic equation. \cite{perel} For a
homogeneous case, this equation is
\begin{equation}
\frac{\partial \mathbf s}{\partial t} =
\boldsymbol{\omega}(\mathbf n)\times \mathbf s + \hat J_0~ \mathbf
s.
 \label{boltz}
\end{equation}
Here $\hat J_0$ is the Boltzmann collision integral and $\mathbf s =\mathbf s(\mathbf p,t)$ is the
spin density in the momentum space, related to the averaged spin by  $\mathbf S = \int \mathbf s
d^2\mathbf p/(2\pi\hbar)^2.$  We assume that the spin splitting is relatively small: $\omega(\mathbf
n) \tau \ll 1.$ This inequality provides   $\tau \ll \tau_S.$ The relationship between $\tau_S$ and
the inelastic scattering  time $\tau_{\rm in}$ varies with temperature. Here we focus on the case
of low temperatures, assuming that $\tau_{\rm in} \gg \tau_S.$ Then spins with different energies do not
correlate  with each other, and  the solution of Eq.~(\ref{boltz}) at $ t \gg \tau$ yields
$~\mathbf S(t)= (m/2\pi\hbar^2)\int \mathbf s_{\rm i} dE $, where $\mathbf s_{\rm i}=\mathbf s_{\rm
i}(E,t)=e^{-t \hat \Gamma} \mathbf s_{\rm i}(E,0)$, $\mathbf s_{\rm i}(0,E)=\langle \mathbf
s(\mathbf p,0)\rangle$ is the initial spin density averaged over the momentum direction, and $\hat
\Gamma=\hat\tau_S^{-1}$ is the spin relaxation tensor (tensor of  inverse relaxation times)
given by \cite{dyak}
\begin{equation}
 \Gamma_{ik}=[\delta_{ik}\sum_{s,l}\Omega^2_{sl}-
\sum_{l}\Omega_{li} \Omega_{lk}]~\tau/2  . \label{Gamma}
\end{equation}

\begin{figure}
\includegraphics[width=0.6\textwidth]{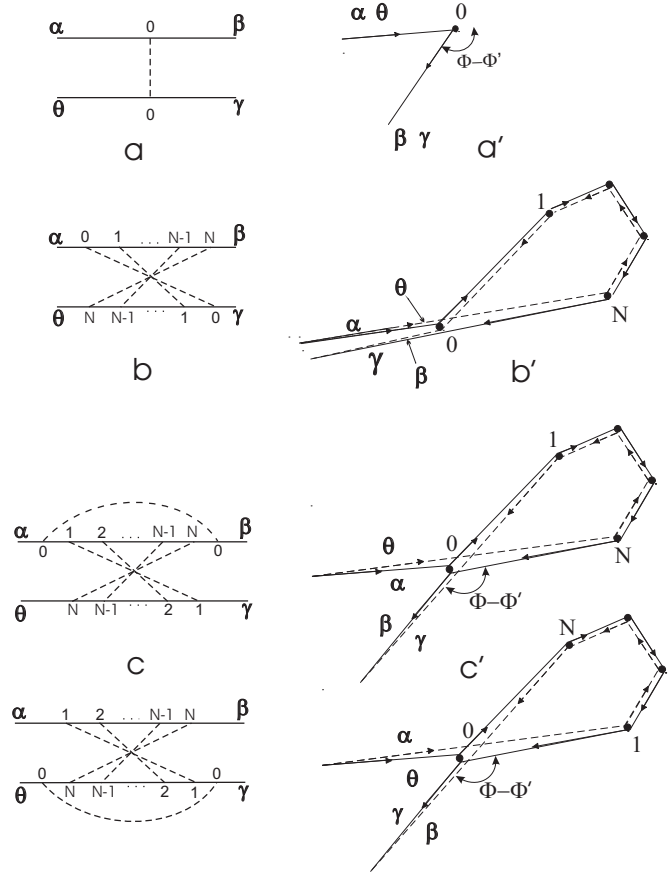}
\caption{ Relevant irreducible diagrams $a,b,c$ and the respective  scattering processes
$a',b',c'$. The Born collision process $a'$ is independent of the electron spin (its contribution
is proportional to $\delta_{\alpha\beta}\delta_{\theta\gamma}$).  Coherent backscattering $b'$
($\Phi-\Phi'\approx \pi$ ), as well as the processes $c'$  describing coherent scattering at an
arbitrary angle ($0<\Phi-\Phi'<2\pi$ ), are spin-dependent due to  rotation of the spin of an
electron passing
along the closed path.} \label{fig1}
\end{figure}

The conventional approach to the calculation of the correlations functions in weakly localized
systems is based on the Kubo formula. \cite{lee,wl} An alternative approach
\cite{ambe,castro,nonbk} is to generalize the Boltzmann equation to include weak localization
effects in the kinetic description. This approach may turn out to be more convenient when
studying   nonlinear and strongly nonequilibrium phenomena. To describe quantitatively the weak
localization phenomenon in the kinetic picture, one has  to modify the Boltzmann equation by
introducing a nonlocal-in-time correction to the collision integral. \cite{ambe,castro} These
corrections can be derived \cite{castro} from the diagrammatic structure of  linear-response
functions. Diagrammatically, the inclusion of a weak localization correction to the effective
collision integral requires the consideration of the irreducible diagrams \cite{castro,nonbk}  in
Fig.~\ref{fig1}b,c,
 in addition to the diagram for the  Born scattering, shown in Fig.~\ref{fig1}a. The crossed-ladder
 diagrams (1b) are usually considered to describe the coherent backscattering of the
electron wave. A physical interpretation of the diagrams (1b,c) in terms of a small change in the
effective differential cross-section for a single impurity
 was suggested in Ref.~\onlinecite{nonbk}.
It was based on an analysis of the interference contribution of  trajectories propagating in the
 opposite directions along closed paths in terms of the phase stationarity
 requirement.
\begin{figure}
\includegraphics[width=0.5\textwidth]{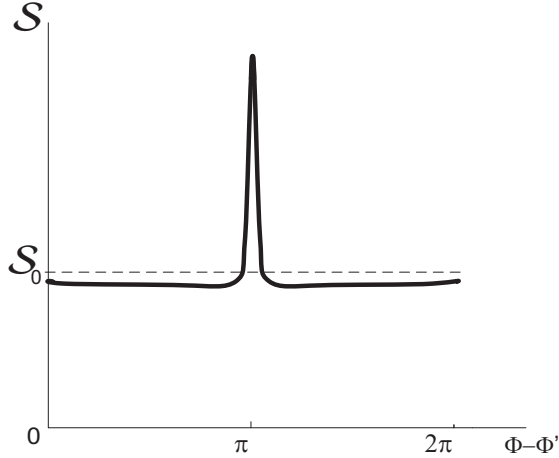}
\caption{ Angular dependence of the effective  cross-section modified by weak localization. The
narrow peak at $\Phi-\Phi'=\pi$ is due to the  coherent backscattering  shown in Fig.~\ref{fig1}b$'$.
The enhancement of the coherent backscattering is accompanied by  a reduction of scattering at
other angles ( the process  in Fig.~\ref{fig1}c$'$), the total cross-section being unchanged.}
\label{fig2}
\end{figure}
 This analysis  shows  that the diagrams (1b)
correspond to a  process shown in Fig.~\ref{fig1}b$'$, which is indeed a  coherent backscattering.
Diagrams (1c) were found  to describe coherent scattering processes with  arbitrary scattering
angles,  shown in Fig.~\ref{fig1}c$'.$

Next, we discuss the key points of a quantitative description of the weakly localized regime within
 the  kinetic approach.  We start with a brief discussion of the zero spin-orbit coupling. As can be seen
 from Figs.~\ref{fig1}b$'$,c$'$, the  relevant processes contain the same closed
paths, so the effective change in the differential cross-section of  impurity $0,$ coming from both
 (1b$'$) and (1c$'$),
is expressed in terms of the return  probability  \cite{nonbk,nonbk1}
\begin{equation}
\frac{\delta\mathcal{S}_{\omega}(\Phi)}{\mathcal{S}_0}= \frac {\lambda l}{\pi}~ W_{\omega}(0)[
\delta(\Phi-\pi)-1/2\pi] . \label{dsdiff}
\end{equation}
  Here $\mathcal{S}_0=1/N_i v \tau$ is the isotropic cross-section in the Drude approximation,
  $N_i$ is the impurity
concentration, $\lambda=2\pi/k$ is the electron wavelength, and $W_{\omega}(0)$ is given by
\begin{equation}
W_{\omega}(0)=\frac{1}{\tau}\int dt e^{i\omega t} W(0 ,t), \label{furww}
\end{equation}
  where
$W(0,t)=W(\mathbf r,t)|_{\mathbf r \to 0}$ is the probability density for a diffusing particle to return
after the time $t$ to the origin $\mathbf r=0.$  The coefficient $\lambda l/\pi$ in the
cross-section correction was found in Ref.~\onlinecite{nonbk} by  integration over small
deviations of the electron trajectories from the trajectories (1b$'$), (1c$'$) meeting the phase
stationarity requirement. Physically, $\lambda l$ is the characteristic  area of the
region around the origin into which the diffusing  electron should return for  the effective
interference to occur. The calculations show that the contributions of (1b$'$) and (1c$'$) have different
signs. The positive contribution represented in Eq.~\eqref{dsdiff} by $\delta(\Phi-\pi)$ comes from
the process (1b$'$), while the negative one (the term $-1/2\pi$), from the process (1c$'$). In other
words, the enhancement of the differential cross-section at the angle $\pi$ due to the coherent
backscattering is accompanied by a reduction of the scattering in other directions, the total
cross-section remaining unchanged (see Fig.~\ref{fig2}). We see that  the correction to the
effective impurity cross-section is $\omega$-dependent. Therefore, the correction to the collision
integral
 in the time representation    turns out to
 be nonlocal in time:
\begin{equation}
\delta \hat J f(\mathbf p, t)= (\lambda l/\pi \tau^2) \int_{-\infty}^t dt' ~W(0,t-t') \int d\Phi'
(\delta(\Phi-\Phi'-\pi)-1/2\pi) f(\mathbf p',t').\label{dJ0}
\end{equation}
Here $f(\mathbf p,t)$ is the electron distribution function (we consider a homogeneous case,
assuming that $f$ is independent of $\mathbf r$) and $\Phi,~\Phi'$ are the angles of $\mathbf p$
and $\mathbf p'.$ Here and further  we omit the "outflux" term  $-(\lambda l/\pi \tau^2) \int_{-\infty}^t dt'
~W(0,t-t') \int d\Phi' (\delta(\Phi-\Phi'-\pi)-1/2\pi) f(\mathbf p,t')$ in the collision integral.
It  makes a zero contribution, because the total cross-section is not changed by the weak
localization.

The  probability density $W_{\omega}(\mathbf r)$ is found as a sum over the   paths involving
different numbers of scattering events (see, for example, Ref.~\onlinecite{chakrob})
\begin{equation}
W_{\omega}(\mathbf r)=\sum_N W^N_{\omega}(\mathbf r), \label{sum-wn}
\end{equation}
where
\begin{equation} W^N_{\omega}(\mathbf r)=\int P(\mathbf r-\mathbf r_N ) P(\mathbf r_N - \mathbf r_{N-1})\dots
P(\mathbf r_2 - \mathbf r_1) P( \mathbf r_1 ) d \mathbf r_1 \dots d \mathbf r_N, \label{wnw}
\end{equation}
 and
\begin{equation}
P(\mathbf r) = \frac{1}{2\pi r l}e^{-r/l+i\omega r/v}. \label{prw}
\end{equation}
At $\omega=0,$ the function $P(r)$ is the classical probability density for an electron starting to move
from
$\mathbf r=0$ to experience the first collision around the point $\mathbf r$. In the framework of a
diagrammatic approach, the function $P(r)$ arises as
 the product of  two  spherical waves
 (retarded and advanced Green's functions) $P(r)=\gamma G^R_{E+\hbar \omega} G^A_E .$
 Here
$G_{R,A}(r) = (\mp i)^{3/2} (m/\hbar^2) e^{ \pm ik r
 -r/2l}
/\sqrt{2 \pi k r}$,  $k=\sqrt{2mE}/\hbar,$ and we took into account that $e^{  ik_{E+\hbar\omega}
r} e^{ - ik_E r} \approx e^{i\omega r/v}.$ A  path involving $N$ scatterings contains $(N+1)$
functions $G_R$ for the clockwise propagation along the path and $(N+1)$ functions $G_A$ for the
counterclockwise propagation. As a result, $W^N_{\omega}$ contains $(N+1)$ functions $P(r).$ Using
Eqs.~\eqref{furww}, \eqref{sum-wn}, \eqref{wnw}, and \eqref{prw}, one can show that in the diffusion
approximation ($\omega \tau \ll 1$), when the typical number of scattering events along a
path is large ($N \gg 1$), the function  $W(\mathbf r,t )$ obeys the diffusion equation
\begin{equation}
\frac{\partial W}{\partial t} - D \Delta W =\delta(\mathbf r)\delta(t). \label{Wdif}
\end{equation}
Solving Eq.~\eqref{Wdif} and taking $\mathbf r=0,$ we find that
the return probability is given by  $W(0,t)=1/4\pi D t  $, and $W_{\omega}(0) \approx
(1/2\pi l^2)\ln(1/\tau \omega).$

A generalization of the above results  to a system with a spin-split spectrum is straightforward.
 Since the electron spin  rotates while passing along a closed loop, electron Green's functions become
 operators with respect to the spin variables:
 \begin{equation}
 \hat G_{R,A}(\mathbf r) = (\mp i)^{3/2} \frac{m}{\hbar^2}\frac{e^{ \pm ik r
 -r/2l}
}{\sqrt{2 \pi k r}}~e^{-i\boldsymbol{\omega}(\mathbf n)\boldsymbol {\sigma}r/2v}. \label{RA}
\end{equation}
They are the products of  spherical waves and  spin-rotation matrices  describing the electron
spin precession.  The  precession frequency $\boldsymbol{\omega}(\mathbf n)$ depends on the
propagation direction $\mathbf n=\mathbf r/r.$  Due to the spin precession,
 the quantum correction to the effective cross section becomes
 spin-dependent: $\delta \mathcal{S}(\Phi) \to \delta\mathcal{S}^{\beta \gamma \alpha  \theta}
 (\Phi),$ where the spin indices   $\alpha, \beta, \gamma, \theta$ correspond
 to the electron trajectories, as shown in Figs.~\ref{fig1}b$'$,c$'$:
\begin{equation}
\frac{\delta \mathcal{S}^{\beta \gamma \alpha \theta}_{\omega}(\Phi)}{\mathcal{S}_0} =\frac{\lambda
l}{\pi}~ W^{\beta\gamma\alpha\theta}_{\omega}(0)[ \delta(\Phi-\pi)-1/2\pi], \label{sabcd}
\end{equation}
The correction to the Boltzmann collision integral can be written as
\begin{equation}
[\delta \hat J \hat f(\mathbf p, \omega)]_{\beta\gamma}=N_iv\int \delta
\mathcal{S}^{\beta\gamma\alpha\theta}_{\omega}(\Phi-\Phi') f_{\alpha\theta}(\mathbf
p',\omega)d\Phi', \label{dJ}
\end{equation}
where $f_{\alpha\theta}(\mathbf p',\omega)$ is the momentum-dependent spin-density matrix. To
derive the expression for $W^{\beta\gamma\alpha\theta}_{\omega}$, we first introduce  the
probability density for a diffusing  electron to arrive after $N$ collisions at the point $\mathbf
r$ with the spin rotated by an angle $\boldsymbol{\phi}$
\begin{equation}
W^N_{\omega}(\mathbf r, \boldsymbol{\phi})=\int
\delta(\boldsymbol{\phi}-\boldsymbol{\phi}_N)P(\mathbf r-\mathbf r_N ) \dots P(\mathbf r_2 -
\mathbf r_1) P( \mathbf r_1 ) d\mathbf r_{1}\dots d\mathbf r_{N}. \label{WN}
\end {equation}
 The  angle
$\boldsymbol{\phi}_N$ changes with the coordinates of the scattering points $\mathbf r_{1},\dots,
\mathbf r_{N}$. One can find it from the matrix equation
\begin{equation}
e^{-i\boldsymbol{\phi}_N\boldsymbol{\sigma}/2} = e^{-i\boldsymbol{\epsilon}(\mathbf r -\mathbf
r_N)\boldsymbol{\sigma}/2} e^{-i\boldsymbol{\epsilon}(\mathbf r_N -\mathbf
r_2)\boldsymbol{\sigma}/2} \dots e^{i\boldsymbol{\epsilon}(\mathbf r_{2}-\mathbf
r_1)\boldsymbol{\sigma}/2} e^{-i\boldsymbol{\epsilon}(\mathbf r_1)\boldsymbol{\sigma}/2}
\label{phi}
\end{equation}
where $\boldsymbol{\epsilon}(\mathbf r)=\boldsymbol{\omega}(\mathbf n)r/v$.
 The spin-dependent return probability is then expressed via the total probability density
\begin{equation}
W_{\omega}(\mathbf r,\boldsymbol{\phi})=\sum_N W_{\omega}^N (\mathbf r,\boldsymbol{\phi})
\label{Wphi0}
\end{equation}
taken at  $\mathbf r=0$
\begin{equation}
W^{\beta\gamma\alpha\theta}_{\omega}(0)=\int
\langle\beta|e^{-i\boldsymbol{\phi}\boldsymbol{\sigma}/2}|\alpha\rangle
\langle\theta|e^{-i\boldsymbol{\phi}\boldsymbol{\sigma}/2}|\gamma\rangle
W_{\omega}(0,\boldsymbol{\phi})~d\boldsymbol{\Lambda}. \label{returnW}
\end{equation}
Here $d\boldsymbol{\Lambda}=g(\phi)d^3\phi$ and function $g(\phi)$ is defined in Appendix
\ref{ApA}. ( Note that, in the absence of the spin-orbit coupling,
$W_{\omega}(0,\boldsymbol{\phi})=\delta(\boldsymbol{\phi})W_{\omega}/g(\phi)$ and
$W^{\beta\gamma\alpha\theta}_{\omega}$ is expressed as $W^{\beta\gamma\alpha\theta}_{\omega} =
\delta_{\alpha\beta}\delta_{\gamma\theta}W_{\omega}$).  What remains to be done is to find an
equation for $W_{\omega}(\mathbf r,\boldsymbol{\phi}).$ To this end,  the probabilities
$W^N_{\omega}$ are related to each other by  the recurrent equations
\begin{equation}
g(\phi)W^{N+1}_{\omega}(\mathbf r, \boldsymbol{\phi})= \int P(\mathbf r')W^N_{\omega}(\mathbf
r-\mathbf r', \boldsymbol{\phi}') \delta(\boldsymbol{\phi}-\boldsymbol{\phi'}-\boldsymbol{\Delta})
d \mathbf r' d\boldsymbol{\Lambda'}. \label{recurrent}
\end{equation}
The vector $\boldsymbol{\Delta}=\boldsymbol{\Delta}_{ \mathbf r', \boldsymbol{\phi'}}$ describes
the change of the spin rotation angle in a ballistic path between two scattering. One can
find it  from the equation
$\exp(-i(\boldsymbol{\phi}'+\boldsymbol{\Delta})\boldsymbol{\sigma}/2)=
\exp(-i\boldsymbol{\epsilon}(\mathbf r')\boldsymbol{\sigma}/2)
\exp(-i\boldsymbol{\phi}'\boldsymbol{\sigma}/2)  .$ Next, we sum Eq.~\eqref{recurrent} over $N$ and
take into account that $\boldsymbol{\Delta} \ll 1.$ After combersome but straightforward calculations, we find
that  $W(\mathbf r,\boldsymbol{\phi},t)=\tau \int W_{\omega}(\mathbf
r,\boldsymbol{\phi})\exp(-i\omega t) d\omega/2\pi $ (the probability density to
arrive after time $t$ to the point $\mathbf r$ with the spin rotated
by the angle $\boldsymbol{\phi}$) is described by  an equation similar to
Eq.~\eqref{Wdif}:
\begin{equation}
\frac{\partial W}{\partial t}- D\left(\frac{\partial}{\partial \mathbf r} - \frac{i\hat \Omega \hat
{\mathbf L}} {v}\right)^2 W =\delta(t) \delta(\mathbf r)\delta(\boldsymbol{\phi})/g(0).
\label{drdif}
\end{equation}
Here $\hat{\mathbf L}$ is the angular momentum operator acting
on the functions of vector $\boldsymbol{\phi}$ (see
Ref.~\onlinecite{edmonds})
and $(\hat \Omega \hat
{\mathbf L})_i = \Omega_{ik} \hat
 L_k.$
 The explicit expressions
for $\hat{\mathbf L}$ and for the common eigen-functions of $\hat L^2$ and $\hat L_z$ are presented
in Appendix \ref{ApA}. Expanding  $\delta(\boldsymbol{\phi})$ in a series over these functions,
 we keep the term with $L=0.$ The
corresponding eigenfunction $\Psi_0$ is independent of the angle $\boldsymbol{\phi}$ and
$\hat{\mathbf L}\Psi_{0}=0$. This is the only term which survives at $t\gg\tau_S.$ The other terms
decay exponentially with  characteristic times of the order of $\tau_S.$ (This statement is not
true for a degenerate case, when $\hat \Omega \hat{\mathbf L}$ depends on the component of
$\hat{\mathbf L}$ along a single axis. This case is discussed below). As a result, we obtain the
following expression for the asymptotical behavior of $W(0,\boldsymbol{\phi},t)$
\begin{equation}
W(0,\boldsymbol{\phi},t)= \frac{1}{4\pi D  t},\quad{\rm for}~~ t\gg \tau_S
 \label{asym}
\end{equation}

Now, we write the distribution function as
\begin{equation}
\hat f = \hat I f + \mathbf s \boldsymbol{\sigma}, \label{hatf}
\end{equation}
where $f$ is the particle density in the momentum space, related to the electron concentration
by $n=2\int f d\mathbf p/(2\pi\hbar)^2,$ $\hat I$ is the unit matrix,  and $\mathbf{s}$ is the spin
density. Substituting Eq.~\eqref{asym} and Eq.~\eqref{hatf} into Eq.~\eqref{returnW} and
Eq.~\eqref{dJ}, respectively, making a Fourier transform, and taking into account Eq.~\eqref{sabcd},
we obtain the
 weak-localization-induced correction to the
collision integral
\begin{align}
&\left(\delta \hat J  f\right)= (\lambda l/\pi \tau^2) \int_{-\infty}^t dt' W(0, t-t')\int
d\Phi'(\delta(\Phi-\Phi'-\pi)-1/2\pi)f(\mathbf p', t'), \label{ro0}
\\
&\left(\delta \hat J \mathbf s\right)_i= (\lambda l/\pi \tau^2)\int_{-\infty}^t dt' W_{ik}(0, t-t')\int
d\Phi'(\delta(\Phi-\Phi'-\pi)-1/2\pi)s_k(\mathbf p', t'), \label{dJ1}
\end{align}
where $W_{ik}(0,t)$ and $W(0,t)$ are given by
\begin{align}
W_{ik}(0, t)= \int
\left(\delta_{ik}-2e_ie_k\sin^2\frac{\phi}{2}\right)~W(0,\boldsymbol{\phi},t)d\boldsymbol{\Lambda},
~~W(0, t)= ~ \int \cos \phi~ W(0,\boldsymbol{\phi},t)d\boldsymbol{\Lambda}. \label{Wik}
\end{align}
Here $\mathbf e =\boldsymbol{\phi}/\phi$.  Using Eq.~\eqref{asym}, we find the asymptotical
behavior of these functions
\begin{equation}
W_{ik}(0,t)= \frac{\delta_{ik}}{8\pi D t}, \quad W(0,t) = - \frac{  1}{8\pi D  t},
\qquad{\rm for} \quad t\gg\tau_S.
 \label{asym1}
\end{equation}
Note also that the spin-orbit coupling  can be neglected in the time interval $\tau \ll t \ll
\tau_S$ . Hence, $W(0,\boldsymbol{\phi},t)\sim \delta(\boldsymbol{\phi})$ and the expressions for
$W_{ik}$ and $W$ become
\begin{equation}
W_{ik}(0,t)= \frac{\delta_{ik}}{4\pi D  t},\quad W(0,t)= \frac{ 1}{4\pi D  t},
\qquad{\rm for}\quad
 \tau \ll t\ll \tau_S.
 \label{asym11}
\end{equation}
We see that the difference between Eqs.~\eqref{asym1} and \eqref{asym11} is in the numerical
coefficients only.
\section*{Solution of the kinetic equation. The long-living tail in the spin polarization}
   For the case with a spin polarization  uniform in space,  the generalized kinetic equation
   is
\begin{equation}
\frac{\partial\mathbf s}{\partial t} = \boldsymbol {\omega} (\mathbf n) \times \mathbf s +
\frac{\mathbf s_{\rm i}-\mathbf s}{\tau} +\delta\hat J~ \mathbf s, \label{boltz10}
\end{equation}
where $\delta\hat J~ \mathbf s$ is given by Eq.~\eqref{dJ1}.  This equation can be solved in the
usual way. \cite{perel} Since $\omega(\mathbf n)\tau \ll 1,$ the spin density can be represented as a
sum
 of the isotropic part $\mathbf s_{\rm i}(E,t),$ which depends on the electron energy only,
and a small anisotropic correction $\mathbf s_{\rm a}(\mathbf p,t),$ which is linear in the
 electron momentum $\mathbf p:$
\begin{equation}
\mathbf s=\mathbf s_{\rm i}+\mathbf s_{\rm a}. \label{s0+s1}
\end{equation}
Substituting  Eq.~\eqref{s0+s1} into Eq.~\eqref{boltz10},  using of Eq.~\eqref{asym1},  and
taking into account the equalities
\begin{equation}
\delta\hat J \mathbf s_{\rm i}=0, \quad \langle \delta\hat J \mathbf s_{\rm a}\rangle=0 , \label{dj}
\end{equation}
(here the angular brackets stand for averaging over the momentum direction) we obtain a closed
relation for $\mathbf s_{\rm i}$
\begin{equation}
\frac{\partial\mathbf s_{\rm i}}{\partial t} = -\hat \Gamma \left( \mathbf s_{\rm i} -
\frac{1}{2\pi kl}\int_{-\infty}^t dt' \frac{\mathbf s_{\rm i}(t')}{t-t'}\right). \label{boltz1}
\end{equation}
Assuming that the spin polarization was created  at $t=0$ with a density $\mathbf s_{\rm
i}(E,0)$ and
neglecting the quantum correction, we get the exponential relaxation $\mathbf s_{\rm i}(E,t)=\theta(t) e^{-\hat
\Gamma t} \mathbf s_{\rm i}(E,0)$ (here $\theta(t)$ is the theta-function). This solution is valid until $e^{-t/\tau_S} \sim 1/\pi k l$. For
larger times, the spin polarization should be found from the condition that  the right-hand side of
Eq.~\eqref{boltz1} equals zero: $\mathbf s_{\rm i}(E,t) \approx (1/2\pi k l)\int_{0}^t dt'
e^{-\Gamma t'}\mathbf s_{\rm i}(E,0)/(t-t').$ So we find that  the spin polarization has a
long-living power tail at large times
\begin{equation}
\mathbf s_{\rm i}(E,t) = \frac{1}{2\pi k l} ~\frac{\hat\Gamma^{-1} }{t} ~ \mathbf s_{\rm i}(E,0).
\label{tail}
\end{equation}
To conclude this section, we note  that we neglected in our calculations  the electron
dephasing due to  inelastic scattering. Such
dephasing  can be accounted for phenomenologically by introducing the factor $\exp(-t/\tau_{\varphi})$
into the right-hand side of Eq.~\eqref{tail}. Here $\tau_{\varphi}$ is the phase-breaking time.
\section*{The degenerate case}
Equation~\eqref{tail} is  invalid for the degenerate case, when the spin precession frequency
$\boldsymbol{\omega}(\mathbf n)$ is parallel to a certain vector $\mathbf u$
 for any  electron momentum: $\boldsymbol{\omega}(\mathbf n) \parallel \mathbf u$
for any $\mathbf n.$  In the classical limit, the component of the electron spin parallel to
$\mathbf u$ does not relax \cite{coment} $\Gamma_{uu}=0$ and the two perpendicular components relax
with equal rates $\Gamma_1=\Gamma_2=\Gamma,$ the off-diagonal components of $\hat\Gamma$ being
equal to zero ($\Gamma_{1 v}=\Gamma_{2v}=\Gamma_{12}=0$ ). This happens in  symmetric quantum wells
grown in the [110] direction, \cite{dyak} as well as   in asymmetric quantum wells grown in the
[001] direction, due to the interplay between Dresselhaus and Rashba couplings. \cite{pikus,golub}
   To find the long-time asymptotic of the return probability, we write the formal solution of
Eq.~\eqref{drdif} as
\begin{equation}
 W(0,\boldsymbol{\phi},t) =\int\frac{d^2 q}{(2\pi)^2}e^{-D(\mathbf q - \hat \Omega
\hat{\mathbf L}/v)^2t} \delta(\boldsymbol{\phi})/g(0) .
 \label{formal}
\end{equation}
In the degenerate case, each of the three operators $(\hat\Omega\hat{\mathbf
L})_x,~(\hat\Omega\hat{\mathbf L})_y$
 and $(\hat\Omega\hat{\mathbf L})_x$ is proportional to $\hat{   L}_u,$ which is the
 component of $\hat{\mathbf L}$ along the precession
 axis. Therefore, these three operators commute with each other. Changing the
 integration variables
 $\mathbf q \to \mathbf q - i\hat \Omega
\hat{\mathbf L}/v$ in Eq.~\eqref{formal}  we obtain
\begin{equation}
W(0,\boldsymbol{\phi},t)=\frac{\delta(\boldsymbol{\phi})}{4\pi g(0)D t}.
 \label{deger}
\end{equation}
Thus, after travelling around a closed loop,  the electron spin does not rotate at all. This can be
interpreted as follows. \cite{pikus}  For the degenerate case, the spin rotation angle is simply
given by $\boldsymbol{\phi}=\int \boldsymbol{\omega}dt \sim \int \mathbf p dt.$  For a closed loop,
we have  $\int \mathbf p dt=0$ and, as a consequence,   $\boldsymbol{\phi}=0.$ After substituting
Eq.~\eqref{deger} into Eq.~\eqref{Wik} and using Eqs.~\eqref{dJ1}, \eqref{boltz10}, one can see that
the weak localization does not affect the longitudinal component of the spin density $ \mathbf u
\mathbf s_{\rm i}( E, t)=\mathbf u \mathbf s_{\rm i}(E,0), $ while the   relaxation of the
perpendicular component is described by the following equation
\begin{equation}
\frac{\partial\mathbf s_{\rm i}}{\partial t} = - \Gamma \left( \mathbf s_{\rm i} - \frac{1}{\pi
kl}\int_{-\infty}^t dt' \frac{\mathbf s_{\rm i}(t')}{t-t'}\right), ~~
 \mathbf s_{\rm i} \perp \mathbf u. \label{deg0}
\end{equation}
 The integral in the right-hand side of Eq.~\eqref{deg0} contains  an additional factor $2$
 as compared  with
 that in Eq.~\eqref{boltz1}. It can be shown that this factor arises from the contribution
 of the eigenfunctions with
 $L=1$ to the long-time asymptotic \eqref{deger}  of the return
 probability. \cite{triplet} The long-time asymptotic of the spin polarization is given by
\begin{equation}
\mathbf s_{\rm i}(E,t) = \frac{1}{\pi k l} ~\frac{1  }{\Gamma t} ~ \mathbf s_{\rm i}(E,0), ~~{\rm
where }~~
 \mathbf s_{\rm i}(E,0) \perp \mathbf u.
\label{degtail}
\end{equation}

\section*{Suppression  of  long-living  polarization by the magnetic field }
Next, we consider the influence of the external magnetic field on the long-living power tail in the spin
polarization.  When an external  field $\mathbf B$ is present, the spin rotation matrices
appearing in
$\hat G_R$ and $\hat G_A$  become different: $e^{-i\boldsymbol{\omega}(\mathbf n)\boldsymbol
{\sigma}r/2v} \to e^{-i\left(\boldsymbol{\omega}(\mathbf n)+\boldsymbol{\Omega}_0\right)\boldsymbol
{\sigma}r/2v} $ for $\hat G_R$ and $e^{-i\boldsymbol{\omega}(\mathbf n)\boldsymbol {\sigma}r/2v}
\to e^{-i(\boldsymbol{\omega}(\mathbf n)-\boldsymbol{\Omega}_0)\boldsymbol {\sigma}r/2v} $ for
$\hat G_A.$  Here, $\boldsymbol{\Omega}_0$ is the frequency of the spin precession in the  field
$\mathbf B$. So, Eq.~\eqref{returnW} is modified as
\begin{equation}
W^{ \beta \gamma \alpha \theta}_{\omega}= \int \langle
\beta|e^{-i\boldsymbol{\phi}\boldsymbol{\sigma}/2}|\alpha\rangle
\langle\theta|e^{-i\boldsymbol{\phi}' \boldsymbol{\sigma}/2}|\gamma\rangle~
W_{\omega}(0,\boldsymbol{\phi},\boldsymbol{\phi}') d \boldsymbol{\Lambda}~d \boldsymbol{\Lambda}',
\label{www}
 \end{equation}
where
\begin{equation}
W_{\omega}(\mathbf r,\boldsymbol{\phi},\boldsymbol{\phi}')= \sum_N \int
\delta(\boldsymbol{\phi}-\boldsymbol{\phi}_N)\delta(\boldsymbol{\phi}'-\boldsymbol{\phi}'_N)P_B(\mathbf
r-\mathbf r_N ) \dots P_B(\mathbf r_2 - \mathbf r_1) P_B( \mathbf r_1 ) d\mathbf r_{1}\dots
d\mathbf r_{N}. \label{w12}
\end{equation}
Here \cite{chakrob}
\begin{equation}
P_B(\mathbf r-\mathbf r')=P(\mathbf r-\mathbf r') \exp\left(i\frac{e\mathbf B_{\perp}}{\hbar c}
[\mathbf r \times \mathbf r'] \right),
\end{equation}
and $\mathbf B_{\perp}$ is the component of the magnetic field normal to the quantum well plane.
 The vectors $\boldsymbol{\phi}_N$ and  $\boldsymbol{\phi}'_N$ should be
 found from Eq.~\eqref{phi}  with
 $\quad \boldsymbol{\epsilon}(\mathbf r)=[\boldsymbol{\omega}(\mathbf
n)+\boldsymbol{\Omega_0}]r/v$  for $\boldsymbol{\phi}_N$ and  $\boldsymbol{\epsilon}'(\mathbf
r)=[\boldsymbol{\omega}(\mathbf n)-\boldsymbol{\Omega_0}]r/v$ for $\boldsymbol{\phi}'_N$. Writing
out the recurrent equations for the consecutive terms in Eq.~\eqref{w12}, we can derive the following
expression
\begin{equation}
\frac{\partial W}{\partial t}-i\boldsymbol{\Omega}_0(\hat{ \mathbf L}-\hat{\mathbf L}')W-
D\left(\frac{\partial}{\partial \mathbf r}-i\frac{2e\mathbf A}{c\hbar} - i\frac{\hat \Omega
(\hat{\mathbf L}+\hat{\mathbf L}')}{v} \right)^2 W =\delta(t) \delta(\mathbf
r)\delta(\boldsymbol{\phi})\delta(\boldsymbol{\phi}')/ g^2(0), \label{WB}
\end{equation}
for the  probability density $W(\mathbf r,\boldsymbol{\phi},\boldsymbol{\phi}',t)=\tau \int
W_{\omega}(\mathbf r,\boldsymbol{\phi},\boldsymbol{\phi}') \exp(-i\omega t)d\omega/2\pi.$  Here the
operators $\hat{\mathbf L}$ and $\hat{\mathbf L}'$ are given by  Eq.~\eqref{L} (with the
replacement  $\boldsymbol{\phi}\to\boldsymbol{\phi}'$ for $\hat{\mathbf L}'$). The term
$-i2e\mathbf A/c\hbar$ in Eq.~\eqref{WB} is responsible for the magnetic field effect on the
electron  orbital motion. \cite{2A} Here $\mathbf A= [\mathbf B_{\perp} \times \mathbf r]/2$   is
the vector potential.
 We assume that the magnetic field is small
\begin{equation}
\Omega_0 \tau_S\ll 1, ~{\rm and}~ DeB_{\perp}\tau_S/\hbar c \ll 1. \label{condit}
\end{equation}

For $\mathbf B=0,$ the solution of  Eq.~\eqref{WB} is
\begin{equation}
W(\mathbf r,\boldsymbol{\phi},\boldsymbol{\phi'},t)= W\left(\mathbf r, \boldsymbol{\phi},t\right)
\delta(\boldsymbol{\phi}-\boldsymbol{\phi'})/g(\phi),
\label{2ugla}
\end{equation}
where $W(\mathbf r,\phi,t)$ obeys Eq.~\eqref{drdif}. Using Eq.~\eqref{2ugla}, we rewrite the
long-living solution \eqref{asym} (which corresponds to zero  total angular momentum
$\hat{\mathbf J} =\hat{\mathbf L} +\hat{\mathbf L}'$) in terms of two angles
$\boldsymbol{\phi},\boldsymbol{\phi'}$:
\begin{equation}
\frac{1}{4\pi D t}\frac{\delta(\boldsymbol{\phi}-\boldsymbol{\phi}')}{ g(\phi)} =\int
\frac{d^2q}{(2\pi)^2}e^{-\gamma(q)t }~\frac{\delta(\boldsymbol{\phi}-\boldsymbol{\phi}')}{g(\phi)},
~~{\rm for} ~~B=0. \label{razlozh-0}
\end{equation}
Here $\gamma(q)=Dq^2$ is the eigenvalue of the operator $\hat \gamma=  D(\mathbf q-\hat \Omega
\hat{ \mathbf J}/v )^2$ at  $J=0$ and $ \delta(\boldsymbol{\phi}-\boldsymbol{\phi}')/g(\phi) $ is
the respective eigenfunction. For $B \neq 0,$ the long time dynamics of the spin relaxation is
determined by the eigenvalues of the operator
\begin{equation}
\hat \gamma=-i\boldsymbol{\Omega}_0(\hat{\mathbf L}-\hat{\mathbf L}')+ D\left(-i\partial/\partial
\mathbf r-2e\mathbf A/c\hbar - \hat \Omega \hat{\mathbf J}/v \right)^2 .
\end{equation}
 First,
we neglect the term $-i\boldsymbol{\Omega}_0(\hat{\mathbf L}-\hat{\mathbf L}').$ In this
approximation,
 the eigenvalues of  $\hat \gamma,$
 corresponding to  $J=0,$ are given by
\begin{equation}
\gamma_n=\gamma_{\perp}(n+1/2), \label{gamman}
\end{equation}
where $\gamma_{\perp}=4eB_{\perp}D/\hbar c$ and $n=0,1,2,\dots.$ Since the operator
$D(-i\partial/\partial \mathbf r-2e\mathbf A/c\hbar)^2$ has a discrete spectrum, we have to make
the replacement
\begin{equation}
\int\frac {d^2q}{(2\pi)^2} e^{-\gamma(q)t}\to \frac{\gamma_{\perp}}{4\pi D}\sum_n e^{-\gamma_n
t}=\frac{\gamma_{\perp}}{4\pi D \sinh(\gamma_{\perp}t/2)}
 \label{discrete}
\end{equation}
    in
Eq.~\eqref{razlozh-0}.  The second step is to take into account  the term
$-i\boldsymbol{\Omega}_0(\hat{\mathbf L}-\hat{\mathbf L}'),$ taking it to be a small perturbation.
In the first order of the perturbation theory, this term leads to the mixing of the eigenfunctions
with $J=1$ to the  eigenfunction with $J=0$. This mixing can be disregarded at $\Omega_0 \ll \Gamma.$
 Corrections to the eigenvalues $\gamma_n$ arise in the second order only. They are calculated
  in Appendix \ref{ApB}.  Using Eq.~\eqref{razlozh1}, we can show that the
     spin polarization dynamics at large times is
described as
\begin{equation}
 \mathbf s_{\rm i}(E,t) = \frac{1}{2\pi k l}~\frac{\gamma_{\perp}
e^{-\boldsymbol{\Omega}_0\hat\Gamma^{-1}\boldsymbol{\Omega}_0t}}
{2\sinh(\gamma_{\perp}t/2)}~~\hat\Gamma^{-1} \mathbf s_{\rm i}(E,0). \label{sB}
\end{equation}
Thus, we see that the magnetic field does suppress the long-living tail in the spin polarization.
\section*{Discussion}
Next we discuss briefly the physical meaning of the results obtained.   Our calculations were based
on the interpretation of the weak localization phenomena in terms of two scattering processes:
coherent backscattering (see Fig.~\ref{fig1}b$'$) and coherent scattering at an arbitrary angle
(see Fig.~\ref{fig1}  $c'$). The existence of the long-living  spin polarization can be explained
as follows. Both coherent scattering processes  were shown to be nonlocal in time. In other words,
the transition of a spin $\mathbf s$ to a spin $\mathbf s',$ caused by  coherent
scattering,
takes a certain time $t,$ which may be relatively long: $t\gg\tau_S.$  The coherent scattering
events do not change the direction of the electron spin. Indeed, as seen from Eq.~\eqref{asym1},
$W_{ik}(0,t) \sim \delta_{ik}$ and, as a consequence, $\mathbf s'
\parallel \mathbf s.$ Therefore, the  electrons involved in such a scattering can keep
memory about the initial spin polarization during the time much longer than $\tau_S.$ The power law
$1/t$ is due to the proportionality of the probability of  coherent scattering to the
probability of  diffusive return.

The collision integral accounting for both  coherent processes does not change the total scattering
cross-section. Therefore,
 \begin{equation}
 \delta \hat J~ f_{\rm i}=0, \quad \delta \hat J~ \mathbf s_{\rm i}=0. \label{renorm0001}
 \end{equation}
Using Eqs.~\eqref{ro0}, \eqref{dJ1}, \eqref{asym1}, and taking into account that $f_{\rm a}(-\mathbf p,\omega)=-f_{\rm a}(\mathbf p,\omega)$
and $\mathbf s_{\rm a}(-\mathbf p,\omega)=-\mathbf s_{\rm a}(\mathbf p,\omega)$, we
obtain
 \begin{equation}
\delta \hat J~ f_{\rm a}(\mathbf p,\omega)=\frac {\ln\left(1/\omega\tau \right)}{2\pi k l
\tau}~f_{\rm a}(\mathbf p,\omega), \quad \delta \hat J~ \mathbf s_{\rm a}(\mathbf p,\omega)=-\frac
{\ln\left(1/\omega\tau \right)}{2\pi k l \tau}~\mathbf s_{\rm a}(\mathbf p,\omega), ~~{\rm
for}~~
 \omega \tau_S\ll 1. \label{renorm}
 \end{equation}
  Eqs.~\eqref{renorm0001}, \eqref{renorm}
imply that the effect of  localization  on the angular spin diffusion, as well as that on the  particle
diffusion,  can be accounted for by the $\omega$-dependent renormalization of the transport
scattering time.  However,
 the quantum corrections to this time   have
differnt signs for the particle  and angular spin diffusion:
\begin{equation}
\frac{1}{\tau_{\rm tr}}=\frac{1- \ln\left(1/\omega\tau \right)/2\pi k l}{\tau} \label{tautr}
\end{equation}
 for the particle diffusion and
 \begin{equation}
\frac{1}{\tau_{\rm tr}'}=\frac{1+\ln\left(1/\omega\tau \right)/2\pi k l}{\tau} \label{tautr1}
\end{equation}
 for the spin diffusion. This implies that the
Eq.~\eqref{tau} is invalid in the quantum case, or, more precisely, it  relates
 $\tau_S$ and $\tau_{\rm tr}'$ rather than  $\tau_S$ and $\tau_{\rm tr}$.

An important comment should be made concerning the role of the coherent nonbackscattering contribution
(processes $c'$ in Fig.~\ref{fig1}).  Neglecting this contribution, we
have
$\delta \hat J \mathbf s_i \neq 0.$ It can be easily shown that this   leads to a physically
meaningless result for the spin relaxation rate.
Therefore, the correct treatment of the effect of weak localization  on the spin relaxation  is only
possible when the nonbackscattering coherent processes are taken into account (the role of such
effects for particle diffusion was discussed in Ref.~\onlinecite{nonbk}). It is worth noting that the
weak localization effects are usually considered to be due to the coherent backscattering only. The
point is that the quantum correction to the conductivity is usually calculated by means of the
Kubo formula, which expresses conductivity in terms of the current-current correlation function.
This approach  focuses on the calculation of the velocity correlation
function, which depends on the anisotropic part of the distribution function $f_{\rm a}$; so, there
is no need to know corrections to $f_i.$ The situation is quite different for spin relaxation which
is due to the relaxation of the isotropic part of the distribution
function. \cite{density}

 To conclude, we have discussed the long-time dynamics of the spin
polarization in a 2D disordered semiconductor. It is shown that, at large times, the spin relaxation
slows down due to weak localization effects. An analytical expression for the long-living tail  of the
spin-polarization has been derived. The magnetic field, both transversal and longitudinal, suppresses
this tail, restoring the exponential relaxation.

 The authors are grateful to A.P. Dmitriev  and K. Kavokin for useful discussions.
 This work has been supported by  RFBR, a grant of the
RAS and a grant of the Russian Scientific School 2192.2003.2 .


\appendix \section{\label{ApA}}
 The analytical
expression for the  angular momentum operator  $\hat{\mathbf L}$ is
\begin{equation}
\hat{\mathbf L} = i\left(\frac{\partial}{\partial \boldsymbol{\phi}}+\left(1-\frac{\phi}{2}
\cot\frac{\phi}{2}\right)\left[ \mathbf e \times \left[\mathbf e \times\frac{\partial}{\partial
\boldsymbol{\phi}}\right]\right]- \left[\frac{\boldsymbol{\phi}}{2}\times\frac{\partial}{\partial
\boldsymbol{\phi}}\right]\right), \label{L}
\end{equation}
with $\mathbf e=\boldsymbol{\phi}/\phi.$
The components of this operator  obey the usual commutation rules,
$[\hat L_i, \hat L_j]=i\epsilon_{ijl}\hat L_l.$
The common eigenfunctions of $\hat L^2$ and $\hat L_z$
are Wigner's rotation matrices \cite{edmonds} $D^L_{MM'}(\boldsymbol{\phi})$:
\begin{equation}\hat L^2
D^L_{MM'}(\boldsymbol{\phi})=L(L+1)D^L_{MM'}(\boldsymbol{\phi}),~~ \hat L_z
D^L_{MM'}(\boldsymbol{\phi})=M D^L_{MM'}(\boldsymbol{\phi}). \label{dlmm}
\end{equation}
 The values of  $L$ may be
integer and half-integer $L=0,1/2,1,3/2, \dots$
  (as usual, $M=-L,\dots,L$). There is also $(2L+1)$ degeneracy with respect to
 $M'.$
The eigenfunction corresponding to $L=0$  is equal to unity
\begin{equation}
\Psi_0=D^0_{00}=1, \quad \hat{\mathbf L}\Psi_0=0.
 \label{psi0}
\end{equation}
The  orthogonality conditions  are \cite{edmonds}
\begin{equation}
\int d\boldsymbol{\Lambda} D^L_{MM'}(\boldsymbol{\phi})D^{L_1*}_{M_1M_1'}(\boldsymbol{\phi})
=\delta_{LL_1}\delta_{MM_1}\delta_{M'M'_1}/(2L+1).
 \label{orto}
 \end{equation}
Here the integration is made over all possible transformations of the $SU(2)$ group. The
integration measure is taken in the invariant form:
\begin{equation}
d\boldsymbol{\Lambda} = ~g(\phi) d^3\boldsymbol{\phi}, \quad{\rm where}\quad
g(\phi)=\frac{1}{16\pi^2}\frac{\sin^2(\phi/2)}{(\phi/2)^2}.
 \label{equation}
 \end{equation}
Note that   $SU(2)$ transformations are usually parameterized by Euler angles
$0<\alpha<2\pi,~0<\beta<\pi,~-2\pi<\gamma<2\pi$ . For such a parametrization, the invariant
integration measure is given by \cite{edmonds} $d\boldsymbol{\Lambda}=\sin\beta ~d\alpha~d\beta
~d\gamma/16\pi^2$ (the expression for $\hat{\mathbf L}$ via the Euler angles is also given in
Ref.~\onlinecite{edmonds}). It is convenient for us to use the components of vector
$\boldsymbol{\phi}$ ($0<\phi<2\pi$) instead of Euler angles. This gives\cite{hamm}
$d\boldsymbol{\Lambda}=g(\phi) d^3\boldsymbol{\phi}. $ The expansion of the delta function
$\delta(\boldsymbol{\phi}-\boldsymbol{\phi'})$ in Wigner's functions is
\begin{equation}
\delta(\boldsymbol{\phi}-\boldsymbol{\phi'})=\sum_{L,M,M'}(2L+1)~D^L_{MM'}(\boldsymbol{\phi})
D^{L*}_{MM'}(\boldsymbol{\phi'})g(\boldsymbol{\phi}).
\label{deltaexpan}
\end{equation}
The solution to Eq.~\eqref{drdif} can be represented as a sum over all angular momenta
$L=0,1/2,1,3/2, \dots .$ However, as follows from Eq.~\eqref{returnW} only two terms ($L=0$ and $L=1$) contribute
to the spin-dependent return probability. Indeed, the action of the operator $\mathbf L$ on the spin-rotation
matrices
is given by
\begin{align}
&\hat{\mathbf L}
\langle\beta|e^{-i\boldsymbol{\phi}\boldsymbol{\sigma}/2}|\alpha\rangle=
\frac{\boldsymbol{\sigma}_{\beta \beta'}}{2}\langle\beta'|e^{-i\boldsymbol{\phi}
\boldsymbol{\sigma}/2}|\alpha\rangle,
\label{iord1} \\
&\hat{\mathbf L}\langle\beta|e^{-i\boldsymbol{\phi}
\boldsymbol{\sigma}/2}|\alpha\rangle\langle\theta|e^{-i\boldsymbol{\phi}\boldsymbol{\sigma}/2}|\gamma\rangle=
\frac{\boldsymbol{\sigma}_{\beta \beta'}\delta_{\theta \theta'} + \delta_{\beta \beta'}\boldsymbol{\sigma}_{\theta \theta'} }{2}\langle\beta'|e^{-i\boldsymbol{\phi}
\boldsymbol{\sigma}/2}|\alpha\rangle\langle\theta'|e^{-i\boldsymbol{\phi}\boldsymbol{\sigma}/2}|\gamma\rangle.
\label{iord2}
\end{align}
As follows from  Eq.~\eqref{iord2}, the projection of the operator $\mathbf L$
onto the subspace formed by products of two rotation matrices  is given by
\begin{equation}
\hat{\mathbf
L}^{\prime}=\frac{\hat{\boldsymbol{\sigma}}^{(1)}+\hat{\boldsymbol{\sigma}}^{(2)}}{2},
\label{sumsigma}
\end{equation}
where $\boldsymbol{\sigma}^{(1)}$ and $\boldsymbol{\sigma}^{(2)}$
are the Pauli matrices acting on the first and second rotation
matrices, respectively. Therefore, the angular momentum
$L'$
can  only be 0 or 1 (singlet and triplet contributions).

Using Eq.~\eqref{drdif} and the property
$\int d\boldsymbol{\Lambda} \Psi_1^* \hat{\mathbf
L}\Psi_2=-\int d\boldsymbol{\Lambda} \Psi_2 \hat{\mathbf
L}\Psi_1^*,$
 valid for arbitrary functions $\Psi_1(\boldsymbol{\phi})$
and $\Psi_2(\boldsymbol{\phi}),$ we can see that the function
\begin{equation}
W^{\beta\gamma\alpha\theta}(\mathbf r,t)=\int
\langle\beta|e^{-i\boldsymbol{\phi}\boldsymbol{\sigma}/2}|\alpha\rangle
\langle\theta|e^{-i\boldsymbol{\phi}\boldsymbol{\sigma}/2}|\gamma\rangle
W(\mathbf r,\boldsymbol{\phi},t)~d\boldsymbol{\Lambda}
\label{returnWr}
\end{equation}
obeys the equation similar to Eq.~\eqref{drdif}
\begin{equation}
\left[ \frac{\partial }{\partial t}- D\left(\frac{\partial}{\partial \mathbf r} + \frac{i\hat \Omega \hat
{\mathbf L}'} {v}\right)^2 \right ]_{\beta \beta' \theta \theta'} W^{\beta'\gamma \alpha \theta'}
=\delta(t) \delta(\mathbf r)\delta_{\alpha \beta} \delta_{\gamma
\theta},
\label{drdif1}
\end{equation}
where matrix elements $\mathbf L^{\prime}_{\beta \beta' \theta \theta'}$ are given
by Eq.~\eqref{iord2}. Note that Eq.~\eqref{drdif1} was used  in Refs.~\onlinecite{pikus,iord,knap}
for calculation of weak localization corrections to conductivity.
The
 alternative derivation of the operator $\hat{\mathbf
L}^{\prime}$ was given in Ref.~\onlinecite{m3}

  \section{\label{ApB}}
Using Eq.~\eqref{deltaexpan}, we  expand Eq.~\eqref{razlozh-0} as
\begin{equation}
 \frac{\delta(\boldsymbol{\phi}-\boldsymbol{\phi}')}{g(\phi)} =
 \sum_{L}\sqrt{2L+1} \sum_{M_1=-L}^L \Psi_{L}^{M_1M_1},
\label{razlozh}
\end{equation}
where
\begin{equation}
\Psi_{L}^{M_1
M_2}(\boldsymbol{\phi},\boldsymbol{\phi}')=\sqrt{2L+1}\sum_{M=-L}^{L}D^L_{MM_1}(\boldsymbol{\phi})
D^{*L}_{MM_2}(\boldsymbol{\phi}') \label{psi-m1m2}
\end{equation}
is the full set of the eigenfunctions ($L=0,1/2,1,3/2,\dots$;~~$M_1=-L,\dots,L$;~~$M_2=-L,\dots,L$)
 for zero  total angular momentum $\hat{ \mathbf  J} \Psi_{L}^{M_1 M_2}=0.$
 In the second
order of the perturbation theory, the term $-i\boldsymbol{\Omega}_0(\hat{\mathbf L}-\hat{\mathbf
L}')$ leads to the  corrections to the eigenvalues of the operator $\hat \gamma$. These corrections
are
 different for the functions $\Psi_{L}^{M_1M_1}$ with different values of $L.$  They  are
 calculated as
 \begin{equation}
  D^{-1}\left\langle \Psi_{L}^{M_1M_1}\left\vert\boldsymbol{\Omega}_0(\hat{\mathbf L}-\hat{\mathbf L}')
  \left(
  \hat \Omega \hat{\mathbf J}/v \right)^{-2} ~\boldsymbol{\Omega}_0(\hat{\mathbf L}-\hat{\mathbf L}')\right\vert
\Psi_{L}^{M_1'M_1'}\right\rangle = \Delta \gamma_L \delta_{M_1M_1'}.
 \label{corrections}
 \end{equation}
   Next, we diagonalize    the operator
   $(\hat \Omega \hat{\mathbf J}/v )^{2}$ in the  subspace formed by the three
   functions $(\hat L_x-\hat L_x')\Psi_{L}^{M_1'M_1'},~(\hat L_y-\hat L_y')
   \Psi_{L}^{M_1'M_1'}$, and $(\hat L_z-\hat L_z')\Psi_{L}^{M_1'M_1'}.$
   Direct calculation gives
\begin{equation}
\Delta \gamma_L =\frac{4L(L+1)}{3} \boldsymbol{\Omega}_0 \hat \Gamma^{-1} \boldsymbol{\Omega}_0.
\label{corr}
\end{equation}
The expression for the long-living solution becomes
\begin{equation}
 W(0,\boldsymbol{\phi},\boldsymbol{\phi'},t)\approx \frac{1}{4\pi D}\frac{\gamma_{\perp}/2}{\sinh(\gamma_{\perp}t/2)}
 \sum_{L}\sqrt{2L+1}~e^{-\Delta\gamma_Lt} \sum_{M_1=-L}^L \Psi_{L}^{M_1M_1},
\label{razlozh1}
\end{equation}
 where $~L=0,1/2,1,3/2,\dots$.
 Substituting Eq.~\eqref{razlozh1} into Eq.~\eqref{www} and using Eqs.~\eqref{sabcd}, \eqref{dJ},
and \eqref{hatf} we derive Eq.~\eqref{sB}.

\begin{thebibliography}{99}
\bibitem{nonbk1} Eq.~\eqref{dsdiff} is valid in the diffusion approximation  ($\omega \tau
\ll 1$) when the typical number of scattering events along the closed path is large. In the
ballistic case ($\omega \tau \gtrsim 1$) the change of the effective cross-section can also be
expressed via return probability. \cite{nonbk}  However,  one should take into account the
dependence of the return probability on $\Phi.$
\bibitem{triplet} In the conventional terminology of the weak
localization, this means that in the degenerate case the triplet channel makes the same
contribution as the singlet one, \cite{pikus} while in the non-degenarate case the non-zero
contribution comes from the singlet channel only.
\bibitem{density} Note that the account of diagrams (1c) is also necessary
for calculation of weak localization correction to the density-density correlation function
(see  Ref.~\onlinecite{m1}) .
\end{thebibliography}
\end{document}